\documentclass[5p,numbers]{elsarticle}

\usepackage{amsmath,amssymb,bm}
\usepackage{graphicx}
\usepackage[dvipsnames]{xcolor}
\usepackage{physics}
\usepackage{siunitx}
\usepackage{hyperref}
\usepackage{cleveref}
\biboptions{compress}

\newcommand{\FeO}{\ensuremath{\gamma\text{-Fe}_2\text{O}_3}}

\newcommand{\beq}{\begin{equation}}
\newcommand{\eeq}{\end{equation}}
\newcommand{\bea}{\begin{eqnarray}}
\newcommand{\eea}{\end{eqnarray}}

\journal{Journal of Magnetism and Magnetic Materials}

\begin{document}

\begin{frontmatter}
\title{Effect of spin disorder on the specific loss power of a nanomagnet}

\author{A.~Michels$^{1}$ and H.~Kachkachi$^{2}$\corref{cor1}}
\ead{hamid.kachkachi@univ-perp.fr}
\cortext[cor1]{Corresponding author.}
%
\address{$^{1}$Department of Physics and Materials Science, University of Luxembourg, 162A~avenue de la Faiencerie, L-1511 Luxembourg, Grand Duchy of Luxembourg}
\address{$^{2}$Laboratoire PROMES CNRS (UPR-8521), Universit\'{e} de Perpignan Via Domitia, Rambla de la Thermodynamique, Tecnosud, 66100 Perpignan, France}

\begin{abstract}
Spin non-collinearities in magnetic nanostructures arise from a variety of sources, including structural defects, finite-size effects, boundary or surface effects, Dzyaloshinskii-Moriya exchange coupling, and magnetic vortex formation. While strong forms of spin disorder generally require a numerical treatment, relatively weak non-collinearities induced by surface anisotropy are amenable to the analytical framework of the effective one-spin problem (EOSP). In this work, we exploit this framework to present a qualitative, semi-analytical study of the effect of spin disorder on the specific loss power (SLP) of a single nanomagnet within linear-response theory. Surface-induced spin misalignment mainly manifests as an additional quartic (cubic-symmetry) contribution to the anisotropy energy, parametrized by the ratio $\zeta \equiv K_4/K_2$. We derive a semi-analytical expression for the SLP as a function of $\zeta$ by combining the $\zeta$-dependent equilibrium susceptibility and the relaxation rate obtained within Langer's approach. Our results show that, for systems in the slow-relaxation regime, the SLP is enhanced by spin misalignment, predominantly through the increase of the relaxation rate caused by the lowering of the effective energy barrier. Retaining the full Debye factor reveals that for moderate reduced barriers $\sigma$, where the system is close to the superparamagnetic regime, the SLP can actually \emph{decrease} with increasing spin disorder. The enhancement is asymmetric with respect to the sign of $\zeta$ and depends on the nanomagnet shape (sphere versus cube) through the geometric prefactors in the EOSP mapping.
\end{abstract}


\end{frontmatter}

\section{Introduction}
\label{sec:intro}

Magnetic hyperthermia, the controlled heating of
tissue by magnetic nanoparticles (NPs) subjected to an
alternating magnetic field, is one of the most
promising biomedical applications of nanomagnetism~\cite{Carrey_JAP2011,
Mehdaoui_AFM2011, Haase_Nowak_PRB85_2012, martinez2013learning,
condeetal15jpcc, Ruta_ScientificReports_2015}.
The efficiency of the heating is quantified by the
specific loss power (SLP), defined as the electromagnetic
power absorbed per unit mass of magnetic material.
Within the linear-response theory (LRT), the SLP is
directly proportional to the imaginary part of the
AC susceptibility~\cite{rosensweig02j3m, Hergt_etal_JPMC2006, barreraetal24pra, valdesetal25acsanm, maniotisetal25nanoscaleadv},
\beq
\label{eq:SLP_def}
\mathrm{SLP}
  = \frac{\mu_0\,\omega}{2\pi}\,H_0^2\,\chi''(\omega),
\eeq
where $\omega = 2\pi f$ is the angular frequency of the AC field of amplitude $H_0$.

Extensive theoretical work has been devoted to the
dependence of the SLP on interparticle dipolar
interactions (DI) and on an external DC bias field~\cite{Haase_Nowak_PRB85_2012, Ruta_ScientificReports_2015, dejardin_etal_SAR_JAP_2017, Landi_JMMM_2017}. However, real nanomagnets are not perfect macrospins:
spin non-collinearities are ubiquitous and can be
induced by structural defects, finite-size effects,
and boundary or surface effects~\cite{dimwys94prb, kodber99prb, kacgar01physa291,
kacgar01epjb, igllab01prb, kacdim02prb, kacgar05springer, yanesetal07prb, garkac03prl,
kachkachi07j3m, garanin18prb, lappas2019}.
In fact, the disruption of the spin ordering can be significantly
more severe when caused by magnetic vortex formation~\cite{shinjoetal00science,
usov2018,eliss2026}, spin frustration at
interfaces, or antisymmetric Dzyaloshinskii-Moriya (DM) exchange
coupling~\cite{dzyaloshinsky58jpcs, moriya60pr}.
In such cases, the spin texture may be far from
a simple canted configuration and a full numerical
treatment of the many-spin problem is generally
required~\cite{IglesiasKachkachi_spring2021}.
The influence of spin disorder on the SLP
has received comparatively little attention, despite the
fact that non-collinearities are known to strongly modify
the effective anisotropy of nanomagnets and, therefore,
their magnetization dynamics.
A recent step in this direction is the work of
Fa\'\i lde~\textit{et al.}~\cite{faildeetal24nanoscale},
who showed by kinetic Monte Carlo simulations that going
beyond the uniaxial-anisotropy description --- in particular
accounting for cubic anisotropy contributions --- can
significantly affect the predicted heating performance.
A related strand of work addresses the effect of
strong, spatially uncorrelated (random) anisotropy on
the microwave response of macroscopic ferromagnets~\cite{chudgar24prb,garanchud22prb};
this is a physically distinct disorder regime from the
weak, coherent surface-induced anisotropy treated here.

Among the various sources of spin misalignment,
\emph{surface anisotropy} (SA) is particularly relevant
for small magnetic NPs, where the surface-to-volume ratio is
large. Importantly, the relatively weak spin misalignment
produced by SA --- when the surface anisotropy
remains moderate compared to the exchange coupling ---
is amenable to an \emph{analytical} treatment through
the effective one-spin problem (EOSP) framework.
This makes SA the ideal testing ground for a qualitative
study of how spin non-collinearities affect the SLP,
while keeping the calculation fully analytical.

The purpose of the present work is therefore to
present a qualitative, semi-analytical investigation of
the influence of spin disorder on the SLP of an \emph{isolated}
nanomagnet (i.e., neglecting DI), taking
surface-induced non-collinearities as a concrete and
experimentally relevant example.
We combine the analytical results for the
$\zeta$-dependent equilibrium susceptibility and the
relaxation rate derived in Refs.~\cite{vernayetal14prb}
and \cite{dejardin_etal_SAR_JAP_2017} to obtain a semi-analytical
expression for the SLP as a function of $\zeta$,
a parameter that reflects the strength of the surface-induced spin disorder.
We stress that, although the present analysis is
restricted to the weak-disorder regime captured by the EOSP,
the qualitative trends it reveals --- in particular the
competition between barrier lowering and the Debye
dissipation condition --- are expected to persist, and
likely to be amplified, when stronger forms of spin
disorder are present.

\section{Model and formalism}
\label{sec:model}

\subsection{EOSP energy}

It has been shown that, for weak-to-moderate surface disorder, the energy of a many-spin nanomagnet (a crystallite) can be mapped onto the \emph{effective one-spin problem} (EOSP)~\cite{garkac03prl, kacbon06prb, kachkachi07j3m, yanesetal07prb}.
More precisely, the EOSP is obtained from the many-spin problem (MSP) under certain conditions regarding the surface anisotropy, which should not be too strong with respect to the exchange coupling, and the particle size, which should not be too small (see Refs.~\cite{garkac03prl,kacbon06prb,kachkachi07j3m,yanesetal07prb} for details). The EOSP consists of the net magnetic moment
\begin{equation}
\boldsymbol{m}=\frac{\sum_{i}\boldsymbol{s}_{i}}{\|\sum_{i}\boldsymbol{s}_{i}\|}.\label{eq:NetMagMoment}
\end{equation}
with $\|\boldsymbol{s}_{i}\|=1$, evolving in an effective energy potential given by
\begin{equation}
\mathcal{H}_{\mathrm{eff}}\simeq-k_{2}m_{z}^{2}+k_{4}\sum_{\alpha=x,y,z}m_{\alpha}^{4}.\label{eq:EffPotential}
\end{equation}
The values and signs of the (effective) coefficients $k_{2}$ and
$k_{4}$ are functions of the atomistic parameters $J, K_{\mathrm{c}}, K_{\mathrm{s}}$,
in addition to the size and shape of the nanomagnet and its underlying
crystal lattice~\cite{yanesetal07prb}. In the following, we will
use the dimensionless constants defined by $k_{\mathrm{c}}=K_{\mathrm{c}}/J$
and $k_{\mathrm{s}}=K_{\mathrm{s}}/J$, so that $k_{2}$ and $k_{4}$
are dimensionless.

For a nanomagnet cut out of a simple cubic lattice, we have~\cite{garkac03prl,garanin_prb18}
\begin{equation}
k_{2}=k_{\mathrm{c}}\frac{N_{\mathrm{c}}}{\mathcal{N}},\quad k_{4}\simeq\begin{cases}
\kappa\frac{k_{\mathrm{s}}^{2}}{z}, & \mathrm{sphere},\\
\\
\left(1-0.7/\mathcal{N}^{1/3}\right)^{4}\frac{k_{\mathrm{s}}^{2}}{z}, & \mathrm{cube},
\end{cases}\label{eq:k2k4}
\end{equation}
where $z$ is the coordination number, $\kappa$ represents a (dimensionless)
surface integral, and $N_{\mathrm{c}}$ is the number of atoms in
the core of the nanomagnet (with full coordination), while $\mathcal{N}$
is the total number of atoms (including both the core and surface).
Note that $\kappa$ was derived in Ref.~\cite{garkac03prl}
in the absence of core anisotropy. As shown in Ref.~\cite{AdamsEtal_ps23},
in the presence of the latter, the spin misalignment caused by surface
anisotropy does not propagate to the center of the nanomagnet; it
is ``fended off'' by the uniaxial anisotropy in the core which tends
to align all spins together. The result of this competition is that
$k_{4}$ (or $\kappa$) is multiplied by the factor $N_{\mathrm{c}}/\mathcal{N}$.
Likewise, for both the cube and sphere, $k_{2}$ may be approximated~\cite{kacbon06prb}
by the first expression in Eq.~(\ref{eq:k2k4}).

We define the two dimensionless relevant parameters
\beq
\label{eq:dimless}
\sigma \equiv \frac{K_2 V}{k_BT}, \qquad
\zeta \equiv \frac{K_4}{K_2},
\eeq
so that $\sigma$ is the reduced uniaxial
anisotropy-energy barrier and $\zeta$ measures the
relative strength of the surface-induced
quartic anisotropy.

The sign and magnitude of $\zeta$ depend on the crystal
structure, shape, and surface arrangement of the
nanomagnet, as was studied in detail by Yanes~\textit{et al.}~\cite{yanesetal07prb}.
For particles cut from a simple cubic (sc) lattice, the
effective cubic constant $K_4$ is negative (i.e.,
$\zeta < 0$), favoring alignment of the magnetization
along the cube edges. In contrast, for particles cut from a face-centered
cubic (fcc) lattice, $K_4$ is positive ($\zeta > 0$),
favoring alignment along the cube diagonals~\cite{yanesetal07prb}.
In both cases, $K_4 \propto K_{\mathrm{s}}^2$, so $\zeta$ is
quadratic in the surface anisotropy constant.
For elongated particles, there is an additional
first-order contribution to the effective uniaxial
anisotropy ($K_2$) that is linear in $K_{\mathrm{s}}$ and scales
with the surface-to-volume ratio, but the quartic
contribution remains proportional to $K_{\mathrm{s}}^2$~\cite{yanesetal07prb,garanin18prb}.
The EOSP mapping is valid as long as $|\zeta|$ remains
below a critical value, estimated numerically to be
$\sim 0.25$ for sc and $\sim 0.35$ for fcc lattices~\cite{yanesetal07prb}.

\subsection{AC susceptibility within the LRT}

In the Debye model, the AC susceptibility of a
nanomagnet with oriented uniaxial anisotropy in a
longitudinal probing field reads~\cite{gitetal74prb,shlste93jmmm,garpal00acp}
\beq
\label{eq:chi_debye}
\chi(\omega) = \frac{\chi_{\text{eq}}}{1 + i\omega/\Gamma_0},
\eeq
where $\chi_{\text{eq}}$ is the equilibrium
(static) susceptibility and $\Gamma_0$ is the
longitudinal relaxation rate.
The imaginary part of $\chi$ is then
\beq
\label{eq:chi_imag}
\chi'' = \chi_{\text{eq}}\,\frac{\eta_0}{1+\eta_0^2},
\eeq
with $\eta_0 \equiv \omega/\Gamma_0$.
Using Eq.~(\ref{eq:SLP_def}), the SLP is
\beq
\label{eq:SLP}
\mathrm{SLP}
  = \frac{\mu_0\,\omega}{2\pi}\,H_0^2\,
    \chi_{\text{eq}}\,\frac{\eta_0}{1+\eta_0^2}.
\eeq
We now express $\chi_{\text{eq}}$ and $\Gamma_0$
as functions of $\zeta$.

\subsubsection{Equilibrium susceptibility}

For a noninteracting nanomagnet described by the EOSP
model in zero DC field, the equilibrium susceptibility
in the limit $\sigma \gg 1$ reads~\cite{vernayetal14prb,sabsabietal13prb}
\beq
\label{eq:chi_eq}
\chi_{\text{eq}}(\sigma,\zeta)
  = 2\chi_0^{\perp}\,\sigma\,\chi_{\text{free}}^{(1)},
\eeq
with
$\chi_0^{\perp} \equiv \mu_0 m^2/(2K_2 V)$
and
\beq
\label{eq:chi1free}
\chi_{\text{free}}^{(1)}
  = \left(1-\frac{1}{\sigma}\right)
  + \frac{\zeta}{\sigma}\left(-1+\frac{2}{\sigma}\right).
\eeq
To linear order in $\zeta$, this can be decomposed as
\beq
\label{eq:chi_decomp}
\chi_{\text{eq}} = \chi_{\text{eq}}^{(0)}
  + \zeta\,\delta\chi_{\text{eq}},
\eeq
where $\chi_{\text{eq}}^{(0)} = 2\chi_0^{\perp}(\sigma-1)$
is the standard uniaxial result and
$\delta\chi_{\text{eq}} = 2\chi_0^{\perp}(2/\sigma - 1)$
is the SA correction.
Note that $\delta\chi_{\text{eq}} < 0$ for $\sigma > 2$,
confirming that SA \emph{decreases} the equilibrium
susceptibility~\cite{vernayetal14prb}.

\subsubsection{Relaxation rate}

The relaxation rate of the EOSP nanomagnet in
zero field and for $|\zeta| \gtrsim \zeta_{\text{crit}}$
is computed within Langer's approach~\cite{lan68prl,lan69ap,bra94jap,bra94prb, kac03epl, kac04jml,titovetal05prb,dejardinetal08jpd}\footnote{The critical value $\zeta_{\text{crit}}$ was discussed in Ref.~\cite{vernayetal14prb}.}
In the EOSP potential (\ref{eq:EffPotential}), the cubic
anisotropy breaks the continuous rotational symmetry
around the easy axis and creates four saddle points at
the equator.
For $\zeta > 0$ and $h=0$, the saddle points are
located at $\theta_{\mathrm{s}} = \pi/2$ with
$\varphi_{\mathrm{s}} = \pi/4 + n\pi/2$ ($n=0,1,2,3$),
while for $\zeta < 0$ they lie at
$\varphi_{\mathrm{s}} = n\pi/2$~\cite{vernayetal14prb}.
The overall topology of the energy landscape remains
similar for both signs of $\zeta$; what changes is
the azimuthal locus of the saddle points and the
values of the energy and curvatures thereat.
Langer's formula reads~\cite{vernayetal14prb,dejardinetal08jpd}
\beq
\label{eq:Gamma_Langer}
\tau_{\mathrm{D}} \, \Gamma_0
  = 4 \, \frac{|\nu|}{2\pi} \,
    \frac{2\sigma(1-\zeta)\,\sin\theta_{\mathrm{s}}}
         {\sqrt{|\mu_1^{(\mathrm{s})}\mu_2^{(\mathrm{s})}|}}\,
    e^{\Delta E^{(0)}},
\eeq
where $\tau_{\mathrm{D}}=\left(\lambda\gamma H_{\mathrm{K}}\right)^{-1}$ is the free diffusion time, $H_{\mathrm{K}}=2K_{2}V/M_{\mathrm{s}}$ is the (uniaxial) anisotropy field, and $\gamma \simeq 1.76 \times 10^{11} \, (\mathrm{T s})^{-1}$ the gyromagnetic ratio.

In Eq.~\eqref{eq:Gamma_Langer}, $\nu$ is the attempt frequency determined by the eigenvalues $\mu_{1,2}^{(\mathrm{s})}$ of the energy Hessian at the saddle point~\cite{vernayetal14prb}.
The energy barrier is
\beq
\label{eq:barrier}
-\Delta E^{(0)}
  = \begin{cases}
    \sigma(1 - \zeta/4) & \text{for } \zeta > 0, \\
    \sigma              & \text{for } \zeta < 0.
    \end{cases}
\eeq
For $\zeta > 0$, the barrier is reduced by
$\sigma\zeta/4$ relative to the pure uniaxial value
$\sigma$, leading to an exponential enhancement of the
relaxation rate.
Remarkably, for $\zeta < 0$ the energy barrier remains
equal to $\sigma$ and is thus independent of $\zeta$.
In this case, the SA contribution enters only
through the Kramers prefactor, i.e., through the
eigenvalues $\mu_{1,2}^{(\mathrm{s})}$ and the attempt
frequency $\nu$.
The eigenvalues at the saddle also differ:
for $\zeta > 0$, $\mu_1^{(\mathrm{s})} = \sigma(\zeta+2)$ and
$\mu_2^{(\mathrm{s})} = -2\sigma\zeta$, while for $\zeta < 0$,
$\mu_1^{(\mathrm{s})} = 2\sigma(1+\zeta)$ and
$\mu_2^{(\mathrm{s})} = 2\sigma\zeta$ \cite{vernayetal14prb}.
In the narrow window $|\zeta| < \zeta_{\text{crit}}$,
the saddle-point structure is too flat for
Langer's approach to be valid, and the relaxation
rate reduces to the N\'eel-Brown result~\cite{Neel_jpr1950,brown63pr,Aharoni_PhysRev1969}
$\tau_{\mathrm{D}} \Gamma_{\text{NB}}
 = (2/\sqrt\pi)\,\sigma^{1/2}\,e^{-\sigma}$,
which is independent of $\zeta$.
We note that analytic approximations for
the field-dependent relaxation time of nanoparticle systems
with uniaxial anisotropy have recently been developed
by Davidson~\textit{et al.}~\cite{davidsonetal24prb};
the present work extends the analysis to the case
where a quartic term modifies the energy landscape. The general case with both uniaxial and cubic anisotropy, in the presence of a DC field, is also studied in Ref. \cite{vernayetal14prb}, but here the DC field is ingored for simplicity.

\subsection{SLP as a function of $\zeta$}

Substituting Eqs.~(\ref{eq:chi_eq})--(\ref{eq:Gamma_Langer}) into Eq.~(\ref{eq:SLP}), one obtains a semi-analytical expression for the SLP as a function of $\zeta$, $\sigma$, $\lambda$, and the field parameters $\omega$ and $H_0$.
Note that the Debye factor $\eta_0/(1+\eta_0^2)$ in Eq.~(\ref{eq:SLP}) peaks at $\eta_0 = 1$ and decreases on either side. In the deep slow-relaxation regime $\eta_0 \gg 1$ (large $\sigma$), this factor reduces to $\Gamma_0/\omega$ and the SLP grows with $\Gamma_0$. Conversely, when $\sigma$ is only moderate and the system is close to or below the $\eta_0 = 1$ crossover, increasing $\Gamma_0$ (by increasing $|\zeta|$) can push $\eta_0$ further below unity, causing the SLP to \emph{decrease}, see results in Fig. \ref{fig:SLP}. This nonmonotonic behavior is missed by the slow-relaxation approximation $\text{SLP} \approx (\mu_0/2\pi) \, H_0^2 \, \chi_{\text{eq}} \, \Gamma_0$ and is one of the central results of this work.

Since SA decreases $\chi_{\text{eq}}$ but increases $\Gamma_0$, the net effect on the SLP depends on the competition between these two tendencies, mediated by the position of the system on the Debye curve. For sufficiently large $\sigma$ and $\zeta > 0$, the relaxation rate dominates through barrier lowering [Eq.~(\ref{eq:barrier})], while for $\zeta < 0$ the enhancement enters through the algebraic prefactor.

\section{Results and discussion}
\label{sec:results}

\begin{figure*}[ht!]
\centering
\includegraphics[width=\textwidth]{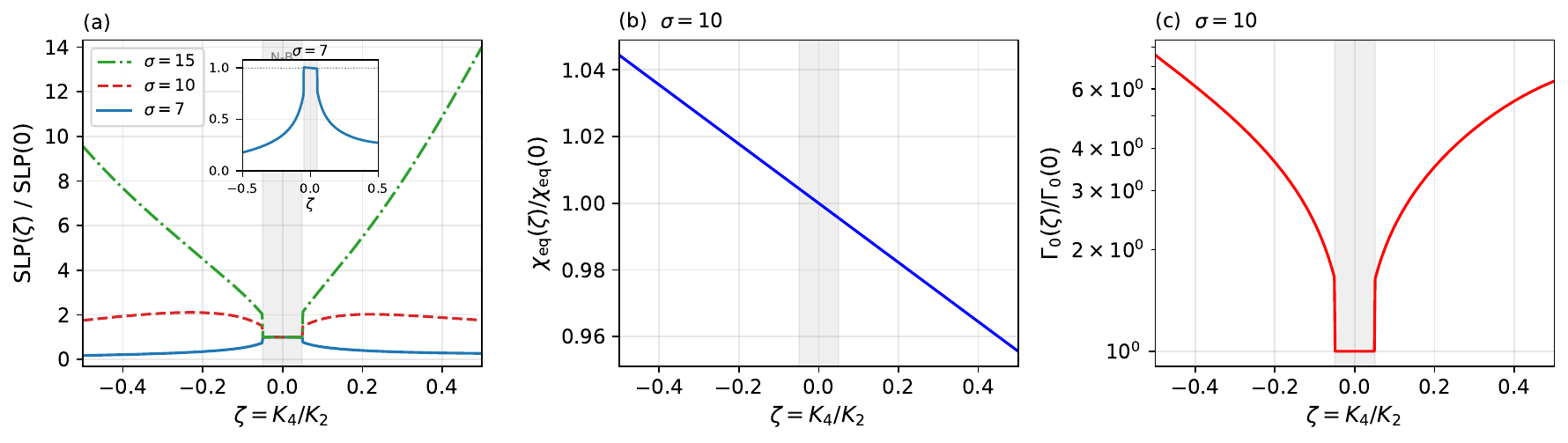}
\caption{%
(a)~SLP normalized to its $\zeta=0$ value as a function
of the surface-anisotropy parameter $\zeta$ for three
values of the reduced barrier $\sigma$.
The inset magnifies the $\sigma = 7$ curve, which
decreases monotonically with $|\zeta|$ (fast-relaxation regime,
$\eta_0 < 1$).
(b)~Equilibrium susceptibility $\chi_{\text{eq}}$,
normalized to its $\zeta=0$ value, for $\sigma = 10$.
(c)~Relaxation rate $\Gamma_0$, normalized to its
$\zeta=0$ value, for $\sigma = 10$.
The shaded band near $\zeta=0$ marks the crossover
region where Langer's approach is not applicable and
the N\'eel-Brown rate is used, which is independent of $\zeta$.
Parameters: $\lambda = 1$, $\tau_0 = 10^{-9}$~s,
$f = 100$~kHz, $\mu_0 H_0 = 10$~mT.}
\label{fig:SLP}
\end{figure*}

We compute the SLP for maghemite ($\FeO$) nanoparticles
of diameter $D = 12$~nm with a saturation magnetization of
$M_{\mathrm{s}} = 4.8\times 10^5$~A/m, an effective uniaxial
anisotropy of $K_2 = 3\times 10^4$~J/m$^3$, and at a temperature of $T = 300$~K.
These parameters yield $\sigma \approx 6.6$.
The AC field has a frequency of $f = 100$~kHz and
an amplitude of $\mu_0 H_0 = 10$~mT.
The damping parameter is set to $\lambda = 1$
and the attempt time to $\tau_0 = 10^{-9}$~s.
A key quantity is the reduced frequency at $\zeta = 0$:
for $\sigma = 7$ one finds $\eta_0 \equiv \omega/\Gamma_0 \approx 0.23$
(fast relaxation), while for $\sigma = 10$ and $15$
one has $\eta_0 \approx 3.9$ and $470$, respectively.

Figure~\ref{fig:SLP}(a) displays the SLP,
normalized to its value at $\zeta = 0$, as a
function of $\zeta$ for three values of $\sigma$.
The full Debye factor in Eq.~(\ref{eq:SLP}) is used
throughout.
Several features emerge:

(i)~The behavior of the SLP depends qualitatively
on $\sigma$. For $\sigma = 15$
($\eta_0 \approx 470$ at $\zeta = 0$, deep slow
relaxation), the SLP is dramatically enhanced by
surface anisotropy: at $\zeta = 0.2$ the SLP is
roughly 5.5 times its $\zeta = 0$ value, rising to
a factor of $\sim 14$ at $\zeta = 0.5$.
For $\sigma = 10$ ($\eta_0 \approx 3.9$), the
enhancement is more modest, reaching a factor of
$\sim 2$ at $|\zeta| = 0.2$.

(ii)~Strikingly, for $\sigma = 7$ ($\eta_0 \approx 0.23$,
fast-relaxation regime), the SLP \emph{decreases} with
$|\zeta|$. This is because the system already lies below
the peak of the Debye factor ($\eta_0 = 1$):
increasing $\Gamma_0$ by surface anisotropy pushes
$\eta_0$ further below unity, reducing $\chi''$.
This nonmonotonic $\sigma$-dependence is entirely absent
from the slow-relaxation approximation and demonstrates
the importance of retaining the full Debye factor.

(iii)~The enhancement is \emph{asymmetric} with
respect to the sign of $\zeta$.
This asymmetry originates from the qualitatively
different mechanisms at play for $\zeta > 0$ versus
$\zeta < 0$ [Eq.~(\ref{eq:barrier})].
For $\zeta > 0$, which according to Yanes~\textit{et al.}~\cite{yanesetal07prb}
corresponds to particles with an fcc internal structure (where the
surface-induced cubic anisotropy is positive),
the energy barrier is reduced by $\sigma\zeta/4$,
producing an exponential boost of the relaxation rate.
For $\zeta < 0$, corresponding to sc-type particles~\cite{yanesetal07prb},
the energy barrier remains exactly
equal to $\sigma$ and the entire enhancement comes
from the modification of the Kramers prefactor:
the eigenvalues $\mu_{1,2}^{(\mathrm{s})}$ at the saddle and
the attempt frequency $\nu$ change with $\zeta$,
yielding a non-negligible (though algebraic rather
than exponential) increase in $\Gamma_0$.
We recall that the EOSP mapping becomes unreliable
for $|\zeta|$ exceeding $\sim 0.25$ on the sc lattice,
whereas a wider range $|\zeta| \lesssim 0.35$ is
accessible on the fcc lattice~\cite{yanesetal07prb}.

(iv)~For large $\sigma$, the effect is more pronounced
at higher $\sigma$: at $\sigma = 15$ the SLP ratio
exceeds 14 for $\zeta = 0.5$.
This is consistent with the exponential sensitivity
of the relaxation rate to the energy barrier for
$\zeta > 0$, $\Gamma_0 \propto \exp(-\sigma(1-\zeta/4))$:
the amplification scales as $e^{\sigma\zeta/4}$,
which is naturally more dramatic for larger $\sigma$.
For $\zeta < 0$, the $\sigma$-dependence enters
through the prefactor, which is a weaker (algebraic)
effect.

The origin of the enhancement is elucidated in
Figs.~\ref{fig:SLP}(b) and \ref{fig:SLP}(c), which
show the individual $\zeta$-dependence of
$\chi_{\text{eq}}$ and $\Gamma_0$ at $\sigma = 10$.
While $\chi_{\text{eq}}$ decreases monotonically
with $\zeta$ [Fig.~\ref{fig:SLP}(b)], the variation
is only $\sim$$5\,\%$ over the range shown.
In contrast, the relaxation rate
[Fig.~\ref{fig:SLP}(c)] increases by nearly an
order of magnitude.
Hence, for systems in the slow-relaxation regime,
the SLP enhancement is overwhelmingly
driven by the increase in the overbarrier
switching rate.

Let us summarize the physical mechanism:
the quartic anisotropy breaks the continuous
azimuthal symmetry around the easy axis and creates
discrete saddle points at the equator.
For $\zeta > 0$, these additional escape routes lower
the effective barrier, producing an exponential
speedup of the thermally activated reversal.
For $\zeta < 0$, the barrier height is unchanged but
the curvatures at the saddle points are modified,
increasing the attempt frequency and thus enhancing
the reversal rate through the Kramers prefactor.
In the context of hyperthermia, this means that
nanomagnets with stronger surface effects dissipate
energy more efficiently \emph{provided} $\sigma$ is
large enough for the system to reside in the
slow-relaxation regime ($\eta_0 > 1$).
For smaller $\sigma$ --- i.e., near the
superparamagnetic regime --- the additional speedup
of the relaxation can be counterproductive by
pushing the system past the optimal $\eta_0 = 1$
condition. The underlying crystal structure matters: the results
of Yanes~\textit{et al.}~\cite{yanesetal07prb} imply that
NPs with an fcc lattice (positive $\zeta$) benefit
from both barrier lowering and prefactor enhancement,
while sc-lattice NPs (negative $\zeta$) are enhanced
only through the prefactor.

\subsection{Shape dependence: sphere versus cube}
\label{sec:shape}

\begin{figure}[t!]
\centering
\includegraphics[width=\columnwidth]{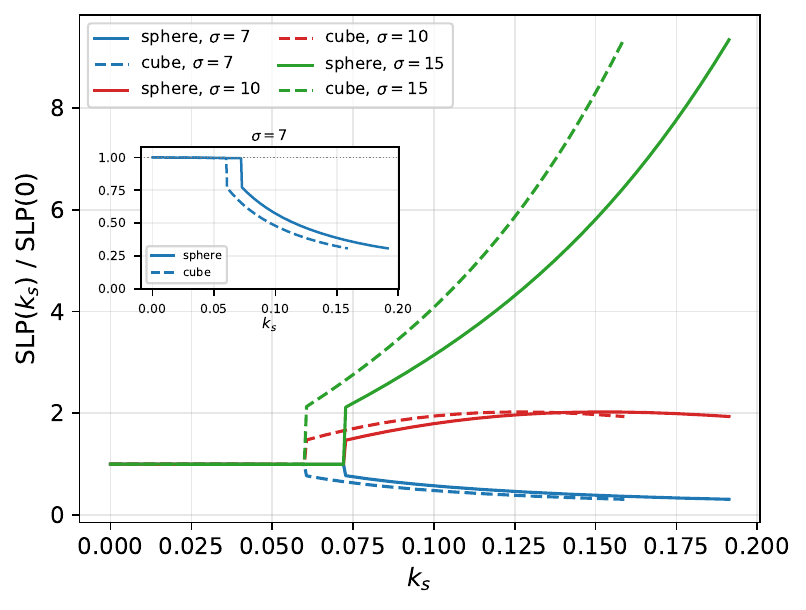}
\caption{%
Normalized SLP as a function of the dimensionless
surface-anisotropy constant $K_{\mathrm{s}}$ for spherical (solid)
and cubic (dashed) nanomagnets on an fcc lattice
($\zeta > 0$), for three values of $\sigma$.
The inset magnifies the $\sigma = 7$ curves, showing
the monotonic SLP decrease characteristic of the
fast-relaxation regime.
The effective cubic coefficient $k_4(K_{\mathrm{s}})$ is computed
from Eq.~(\ref{eq:k2k4}): $k_4 = \kappa\,K_{\mathrm{s}}^2/z$ for
the sphere (with $\kappa = 0.535$) and
$k_4 = (1-0.7/\mathcal{N}^{1/3})^4\,K_{\mathrm{s}}^2/z$ for
the cube.
Parameters: $\mathcal{N} = 1500$, $z = 12$ (fcc),
$k_\mathrm{c} = 0.01$, $N_\mathrm{c}/\mathcal{N} \approx 0.47$.
Curves are clipped at $\zeta = 0.35$ (EOSP validity limit).}
\label{fig:shape}
\end{figure}

The EOSP coefficient $k_4$ depends on the nanomagnet
shape through Eq.~(\ref{eq:k2k4}).
For a given value of the atomistic surface-anisotropy
constant $K_{\mathrm{s}}$, the cube yields a larger effective
$\zeta$ than the sphere because
$(1-0.7/\mathcal{N}^{1/3})^4 > \kappa$ for the
particle sizes of interest.
With $\mathcal{N} = 1500$ on an fcc lattice ($z = 12$)
and $\kappa = 0.535$, the cube-to-sphere ratio is
$(1-0.7/\mathcal{N}^{1/3})^4/\kappa \approx 1.45$.

Figure~\ref{fig:shape} displays the normalized SLP as a
function of $K_{\mathrm{s}}$ for both shapes and several values
of $\sigma$.
For $\sigma = 15$, the cube reaches an SLP enhancement
of nearly an order of magnitude at
$K_{\mathrm{s}} \approx 0.15$, whereas the sphere achieves the same
level of enhancement only at larger $K_{\mathrm{s}}$.
This difference is purely geometric in origin:
the surface integral $\kappa$ for the sphere is
smaller than the corresponding prefactor for the cube,
so that the cubic contribution to the effective
anisotropy builds up more slowly with $K_{\mathrm{s}}$.
The cube also reaches the EOSP validity limit
($\zeta = 0.35$) at a smaller $K_{\mathrm{s}}$ ($\approx 0.16$)
than the sphere ($\approx 0.19$).
For $\sigma = 7$, both shapes show a decrease of the SLP
with $K_{\mathrm{s}}$, confirming the fast-relaxation suppression
discussed above.

\subsection{Temperature dependence of the SLP}
\label{sec:temperature}

\begin{figure}[t!]
\centering
\includegraphics[width=\columnwidth]{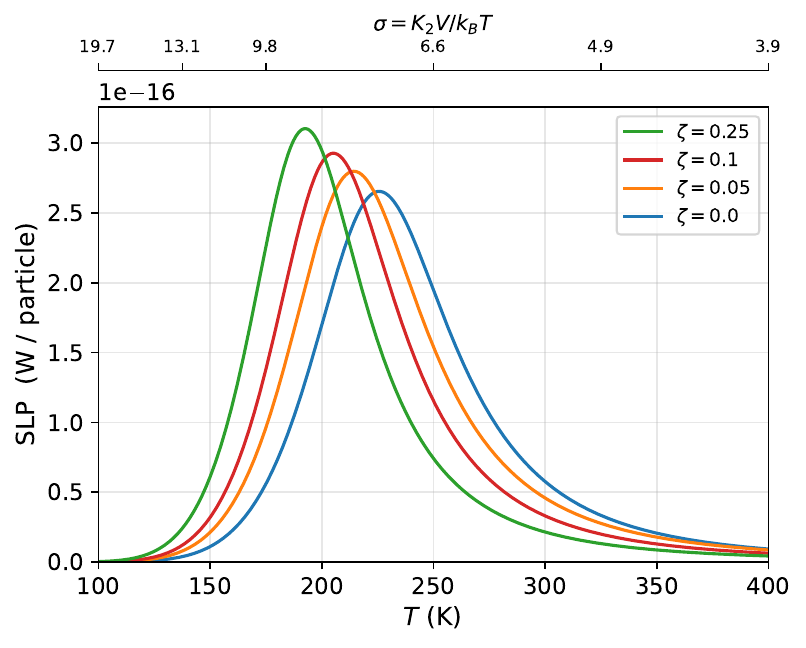}
\caption{%
SLP of a maghemite nanoparticle ($D = 12$~nm) as a
function of temperature for several values of $\zeta$ (see inset).
The top axis shows the corresponding reduced barrier
$\sigma = K_2 V/k_\mathrm{B} T$.
Parameters: $\lambda = 1$, $\tau_0 = 10^{-9}$~s,
$f = 100$~kHz, $\mu_0 H_0 = 10$~mT.}
\label{fig:SLP_T}
\end{figure}

Finally, we examine the temperature dependence of the
SLP at fixed $\zeta$.
Since $\sigma = K_2 V/(k_\mathrm{B} T)$, varying $T$ scans through both the slow- and fast-relaxation regimes.
The SLP is expected to exhibit a maximum at the temperature
$T^{*}$ where $\eta_0 = \omega/\Gamma_0 \approx 1$, i.e.,
where the Debye dissipation is maximal.

Figure~\ref{fig:SLP_T} shows the SLP as a function
of $T$ for $\zeta = 0$, $0.05$, $0.1$, and $0.25$.
All curves display the expected bell shape.
The peak shifts to lower temperature with increasing
$\zeta$: from $T^* \approx 225$~K ($\sigma \approx 8.7$)
at $\zeta = 0$ to $T^* \approx 192$~K
($\sigma \approx 10.2$) at $\zeta = 0.25$.
This shift is a direct consequence of the enhancement
of $\Gamma_0$ by surface anisotropy, which pushes the
$\eta_0 = 1$ crossover to higher $\sigma$
(lower $T$).
The peak amplitude increases modestly ($\sim 17\,\%$
over the range of $\zeta$ shown), because at the
Debye maximum the SLP reduces to
$(\mu_0 \omega/2\pi)\,H_0^2\,\chi_{\text{eq}}/2$,
which depends only weakly on $\zeta$ through
$\chi_{\text{eq}}$.
At temperatures above $T^*$, the system enters the
fast-relaxation regime ($\eta_0 < 1$) and the SLP
decreases; this is the regime in which further
increasing $\Gamma_0$ through surface anisotropy
becomes counterproductive, as discussed in
Sec.~\ref{sec:results}.

\section{Conclusion}
\label{sec:conclusion}

We have presented a qualitative, semi-analytical study of
how spin non-collinearities in nanomagnets affect
the specific loss power (SLP) within the linear-response theory.
By exploiting the analytical framework of the effective one-spin problem (EOSP) and
focusing on the experimentally important case of
surface-induced spin disorder, we have derived a
semi-analytical expression for the SLP as a function of
the surface-anisotropy parameter $\zeta = K_4/K_2$, the
reduced energy barrier $\sigma$, and the damping parameter
$\lambda$. A central result is the importance of the behavior of
the Debye factor $\eta_0/(1+\eta_0^2)$: for systems in the slow-relaxation regime
($\eta_0 \gg 1$), the dominant mechanism is the
increase of the relaxation rate $\Gamma_0$ caused by
the creation of additional saddle points in the energy
landscape, leading to an SLP enhancement.
In contrast, for moderate $\sigma$ values where
$\eta_0 \lesssim 1$ (near the superparamagnetic
regime), the same increase in $\Gamma_0$ can push the
system past the optimal Debye condition, resulting in
a \emph{decrease} of the SLP.
The sign of $\zeta$, which is set by the underlying
crystal lattice as established in Ref.~\cite{yanesetal07prb},
leads to quantitatively different barrier reductions
and therefore to an asymmetric SLP enhancement.
We have also shown that, for a given surface-anisotropy
constant $K_{\mathrm{s}}$, the nanomagnet shape (sphere versus cube)
modulates the effective $\zeta$ through the geometric
prefactors in Eq.~(\ref{eq:k2k4}), with cubes producing
a stronger SLP enhancement than spheres.
The temperature dependence of the SLP exhibits a
bell-shaped curve whose peak shifts to lower temperatures
with increasing $\zeta$.

We emphasize that the present analysis is
restricted to relatively weak spin disorder, where the
EOSP mapping onto a macroscopic potential with uniaxial
and quartic terms remains valid.
Stronger disruptions of the spin order --- such as
magnetic vortex states~\cite{shinjoetal00science,usov2018,eliss2026},
spin frustration at interfaces, or spin textures induced
by Dzyaloshinskii-Moriya exchange coupling~\cite{dzyaloshinsky58jpcs, moriya60pr} ---
can produce qualitatively similar but potentially
much larger modifications of the effective energy
landscape. Such situations lie beyond the scope of analytical
approaches and require full numerical simulations of
the many-spin problem~\cite{IglesiasKachkachi_spring2021}.
A conceptually related but distinct problem is posed by
strong, spatially uncorrelated anisotropy disorder in
extended, exchange-coupled ferromagnets, where a scaling
argument links the weak- and strong-disorder regimes and
predicts a microwave-absorption band that saturates with
spin heating at high power or temperature~\cite{chudgar24prb,garanchud22prb}
--- in contrast to the single-particle, coherent EOSP
treatment developed here. We are currently extending the
comparison to dipolar-interacting nanomagnet chains, where
kinetic Monte Carlo simulations can be compared with
analytical predictions.
The qualitative trends identified here --- the
competition between barrier lowering and the Debye
dissipation condition, the asymmetry with respect to the
sign of the induced anisotropy, and the shape dependence ---
are expected to persist and to serve as guidelines
for interpreting more complex numerical studies.


\end{document}